\documentstyle[aps,prl,twocolumn,psfig,floats]{revtex}
\begin{document}
\draft
\wideabs{
\title{Microscopic universality in the spectrum of the lattice Dirac
  operator}
\author{M.E.~Berbenni-Bitsch$^1$, S.~Meyer$^1$, A.~Sch\"afer$^2$,
  J.J.M.~Verbaarschot$^3$, and T.~Wettig$^{4,5}$}
\address{%
  $^1$Fachbereich Physik -- Theoretische Physik, Universit\"at
  Kaiserslautern, D-67663 Kaiserslautern, Germany\\
  $^2$Institut f\"ur Theoretische Physik, 
  Universit\"at Regensburg, D-93040 Regensburg, Germany\\
  $^3$Department of Physics, State University of New York, Stony
  Brook, NY 11794, USA\\
  $^4$Max-Planck-Institut f\"ur Kernphysik, Postfach 103980, D-69029
  Heidelberg, Germany\\
  $^5$Institut f\"ur Theoretische Physik, Technische Universit\"at
  M\"unchen, D-85747 Garching, Germany}
\date{December 2, 1997}
\maketitle
\begin{abstract}
  Large ensembles of complete spectra of the Euclidean Dirac operator
  for staggered fermions are calculated for SU(2) lattice gauge
  theory.  The accumulation of eigenvalues near zero is analyzed as a
  signal of chiral symmetry breaking and compared with parameter-free
  predictions from chiral random matrix theory.  Excellent agreement
  for the distribution of the smallest eigenvalue and the microscopic
  spectral density is found.  This provides direct evidence for the
  conjecture that these quantities are universal functions.
\end{abstract}
}

\narrowtext 

Hadronic properties such as the lightness of the pion masses and the
absence of parity doublets, strongly indicate that chiral symmetry is
broken spontaneously. In QCD, a great deal of insight in such
nonperturbative phenomena has been obtained from extensive lattice QCD
simulations \cite{DeTar,Ukawa}.  This did not go without a significant
amount of effort.  One of the difficulties is that the order parameter
of the chiral phase transition, $\langle\bar\psi\psi\rangle$, can be
obtained only after a complicated limiting procedure: the
thermodynamic limit, the chiral limit, and the continuum limit.
In addition, in the chiral limit it is extremely costly to take
into account the effect of the fermion determinant. Since the fermion
determinant can be expressed as a product over the Dirac eigenvalues,
this alone warrants a detailed study of the QCD Dirac spectrum. 
Moreover, $\langle\bar\psi\psi\rangle$ is directly
related to the QCD Dirac spectrum through the Banks-Casher
relation \cite{Bank80},
\begin{equation}
  \label{eq1}
  \langle\bar\psi\psi\rangle=\lim_{m\rightarrow 0}
  \lim_{V\rightarrow\infty}\frac{\pi}{V}\rho(0).
\end{equation}
Here, $m$ is the quark mass, $V$ is the volume of space-time, and
$\rho(\lambda)=\langle\sum_n\delta(\lambda-\lambda_n)\rangle$ is the
eigenvalue density of the Euclidean Dirac operator,
$iD=i\gamma_\mu\partial_\mu+\gamma_\mu A_\mu$, averaged over gauge
field configurations. We observe that the average position of the
smallest eigenvalues is determined by the chiral condensate. In this
letter we focus on fluctuations of the smallest eigenvalues about
their average position.  It should be clear that such fluctuations
affect the fermion determinant and are important for the understanding
of finite size effects \cite{Gock}.  The hope is that they are given
by universal functions which can be obtained analytically.  This
analytical information could then be used to facilitate extrapolations
to the thermodynamic and chiral limits.

A similar situation arises in mesoscopic physics \cite{HDgang}.  In
these studies, it was shown that for a sufficient amount of disorder,
spectral correlations are universal and can be obtained from a Random
Matrix Theory (RMT) with only the basic symmetries included.  On the
other hand, the average spectral density is non-universal and requires
specific knowledge of the dynamics of the system.  In this letter we
investigate the question whether a similar separation of scales takes
place in QCD. Does the disorder of lattice QCD gauge field
configurations result in universal fluctuations of the small Dirac
eigenvalues?

According to the Banks-Casher relation, the low-lying Dirac
eigenvalues are spaced as $1/V$ for $\langle\bar\psi\psi\rangle \ne
0$. Recent work by Leutwyler and Smilga \cite{Leut92} shows that this
part of the spectrum is related to the pattern of chiral symmetry
breaking by means of a class of sum rules for the inverse Dirac
eigenvalues. It is natural to magnify the spectrum near $\lambda= 0$
by a factor of $V$.  This leads to the intro\-duction of the
microscopic spectral density \cite{Shur93} defined by
\begin{equation}
  \label{eq2}
  \rho_s(z)=\lim_{V\rightarrow\infty}\frac{1}{V\Sigma}
  \rho\left(\frac{z}{V\Sigma}\right) \:,
\end{equation}
where $\Sigma$ is the absolute value of $\langle\bar\psi\psi\rangle$.
Based on the analysis of the Leutwyler-Smilga sum rules, it was
conjectured \cite{Shur93} that this distribution is universal and only
determined by the global symmetries of the QCD partition function, the
number of flavors, and the topological charge.  If that is the case it
can be obtained from a much simpler theory with only the global
symmetries as input. Such a theory is chiral RMT which will be
discussed below. Whether or not QCD is in this universality class is a
dynamical question that can only be answered by lattice QCD
simulations.  The investigation of this question is the main purpose
of this letter.  At this moment it can only be addressed on relatively
small lattices where our results are consistent with zero topological
charge.  The pertinent question of what happens in the continuum limit
has to be postponed to future work. In this limit we expect zero modes
to become important.  $\rho_s(z)$ is then different in different
sectors of topological charge.  However, on present day lattices with
staggered fermions there seems to be no evidence of a ``zero-mode
zone'' \cite{Kogu97}, and the situation is controversial at best.

There are already several pieces of evidence supporting the conjecture
that $\rho_s$ is universal: (1) The moments of $\rho_s$ generate the
Leutwyler-Smilga sum rules \cite{Verb93}.  (2) $\rho_s$ is insensitive
to the probability distribution of the random matrix
\cite{Brez96,Nish96}. (3) Lattice data for the valence quark mass
dependence of the chiral condensate could be understood using the
analytical expression for $\rho_s$ \cite{Chan95,Verb96a}. (4) The
functional form of $\rho_s$ does not change at finite temperature
\cite{Jack96b}.  (5) The analytical result for $\rho_s$ is found in
the Hofstadter model for universal conductance fluctuations
\cite{Slev93}.  (6) For an instanton liquid $\rho_s$ shows good
agreement with the random-matrix result \cite{Verb94b}.  However, a
direct demonstration for lattice QCD was missing.

An analysis of Dirac spectra on the lattice was performed in
Ref.~\cite{Hala95} where it was shown that the spectral fluctuations
in the bulk of the spectrum on the scale of the mean level spacing are
universal and described by RMT.  This showed that the eigenvalues of
the Dirac operator are strongly correlated.  Only few configurations
were available in this study, but spectral ergodicity allowed to
replace the ensemble average by a spectral average.  However, spectral
averaging is not possible for $\rho_s$ since only the first few
eigenvalues contribute.  Therefore, a large number of configurations
is essential.

We briefly summarize the main ingredients of chiral RMT.  In a
random-matrix model, the matrix elements of the operator under
consideration are replaced by the elements of a random matrix with
suitable symmetry properties.  Here, the operator is the Euclidean
Dirac operator $iD$ which is hermitian.  Because $\gamma_5$
anti-commutes with $iD$ the eigenvalues occur in pairs $\pm\lambda$.
In a chiral basis, the random-matrix model has the structure
\cite{Shur93}
\begin{displaymath}
  iD+im \rightarrow \left[\matrix{im&W\cr W^\dagger&im}\right]\:,
\end{displaymath}
where $W$ is a matrix whose entries are independently distributed
random numbers. In full QCD with $N_f$ flavors, the weight function
used in averaging contains the gluonic action in the form
$\exp(-S_{\rm gl})$ and $N_f$ fermion determinants.  In RMT, the
gluonic part of the weight function is replaced by a Gaussian
distribution of the random matrix $W$.  The symmetries of $W$ are
determined by the anti-unitary symmetries of the Dirac operator.
Depending on the number of colors and the representation of the
fermions the matrix $W$ can be real, complex, or quaternion real
\cite{Verb94a}.  The corresponding random-matrix ensembles are called
chiral Gaussian orthogonal (chGOE), unitary (chGUE), and symplectic
(chGSE) ensemble, respectively.  The microscopic spectral density has
been computed analytically for all three ensembles
\cite{Verb93,Verb94d,Naga95}.

We have performed numerical simulations of lattice QCD with staggered
fermions and gauge group SU(2) for couplings $\beta=4/g^2 = 2.0$, 2.2,
and 2.4 on lattices of size $V=L^4$ with $L=8$, 10, and 16. This range
of lattice parameters covers the crossover region from strong to weak
coupling of SU(2) \cite{Creu80}.  The boundary conditions are periodic 
for the gauge fields and periodic in space and anti-periodic in 
Euclidean time for the fermions.  In this work,
we only study the quenched approximation using a hybrid Monte Carlo
algorithm \cite{Meye90}.  This made it possible to generate a large
number of independent configurations (indicated in the figures).  The
analysis of unquenched data with 4 dynamical flavors is in progress.

In SU(2) with staggered fermions, every eigenvalue of $iD$ is twofold
degenerate due to a global charge conjugation symmetry.  In addition,
the squared Dirac operator $-D^2$ couples only even to even and odd to
odd lattice sites, respectively.  Thus, $-D^2$ has $V/2$ distinct
eigenvalues.  We use the Cullum-Willoughby version of the Lanczos
algorithm \cite{Stoe93} to compute the complete eigenvalue spectrum of
the sparse hermitian matrix $-D^2$ in order to avoid numerical
uncertainties for the low-lying eigenvalues.  There exists an
analytical sum rule, ${\rm tr}(-D^2) = 4V$, for the distinct
eigenvalues of $-D^2$ \cite{Kalk95}.  We have checked that this sum
rule is satisfied by our data, the largest relative deviation was
$\sim 10^{-8}$.  We have also made a detailed study to determine the
optimal acceptance rates and trajectory lengths \cite{Berb97}.  The
integrated autocorrelation times are in the range of 1 to 4.  The
chiral condensate was obtained by fitting the spectral density and
extracting $\rho(0)$ and is given in Table~\ref{table1} below.

\begin{figure}[h]
\vspace*{2mm}
\centerline{\psfig{figure=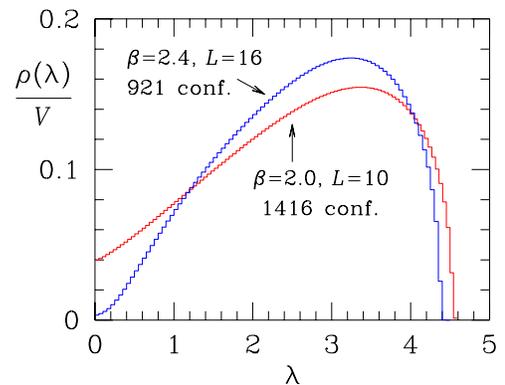,height=50mm}}
\vspace*{2mm}
\caption{Spectral density of the lattice Dirac operator for
  $\beta=2.0$ and 2.4.  Only positive eigenvalues are plotted.}
\label{fig1}
\end{figure}

The overall spectral density of the Dirac operator cannot be obtained
in a random-matrix model since it is not a universal function.  The
lattice result for $\rho(\lambda)$ is displayed in Fig.~\ref{fig1} for
$\beta=2.0$, $V=10^4$ and $\beta=2.4$, $V=16^4$, respectively.  Note
the strong decrease in $\langle\bar\psi\psi\rangle$ (in lattice units)
for $\beta=2.4$, cf.\ Eq.~(\ref{eq1}) and Table~\ref{table1}.

We are particularly interested in the region of small eigenvalues to
check the predictions from chiral RMT.  According to
Ref.~\cite{Verb94a}, staggered fermions in SU(2) have the symmetries
of the chGSE.  Analytical expressions can be obtained in the framework
of RMT for the microscopic spectral density and the distribution of
the smallest eigenvalue by slight modifications of results computed
for Laguerre symplectic ensembles \cite{Naga95,Forr93}.  Incorporating
the chiral structure of the Dirac operator, we obtain from
Ref.~\cite{Naga95}
\begin{eqnarray}
  \label{eq3} 
  \rho_s(z)=2z^2\int_0^1duu^2\int_0^1dv&&[J_{4a-1}(2uvz)J_{4a}(2uz) 
  \nonumber \\  &&-vJ_{4a-1}(2uz)J_{4a}(2uvz)] 
\end{eqnarray}
with $4a=N_f+2\nu+1$, where $N_f$ is the number of massless flavors
and $\nu$ is the topological charge.  For our quenched data, $4a=1$
since $\nu=0$ as explained in the introduction.  According to
Eq.~(\ref{eq2}), lattice data for $\rho_s(z)$ are constructed from the
numerical eigenvalue density using a scale
$V\langle\bar\psi\psi\rangle$.  This scale is determined by the data,
hence the random-matrix predictions are parameter-free.  Similarly,
the distribution of the smallest eigenvalue for $N_f=\nu=0$ follows
from Ref.~\cite{Forr93},
\begin{equation}
  \label{eq4} 
  P(\lambda_{\rm min})=\sqrt{\frac{\pi}{2}}c(c\lambda_{\rm min})^{3/2}
  I_{3/2}(c\lambda_{\rm min})e^{-\frac{1}{2}(c\lambda_{\rm min})^2} \:,
\end{equation}
where $c=V\langle\bar\psi\psi\rangle$ is the same scale as above.
\begin{figure}[t]
\twocolumn[
\centerline{\psfig{figure=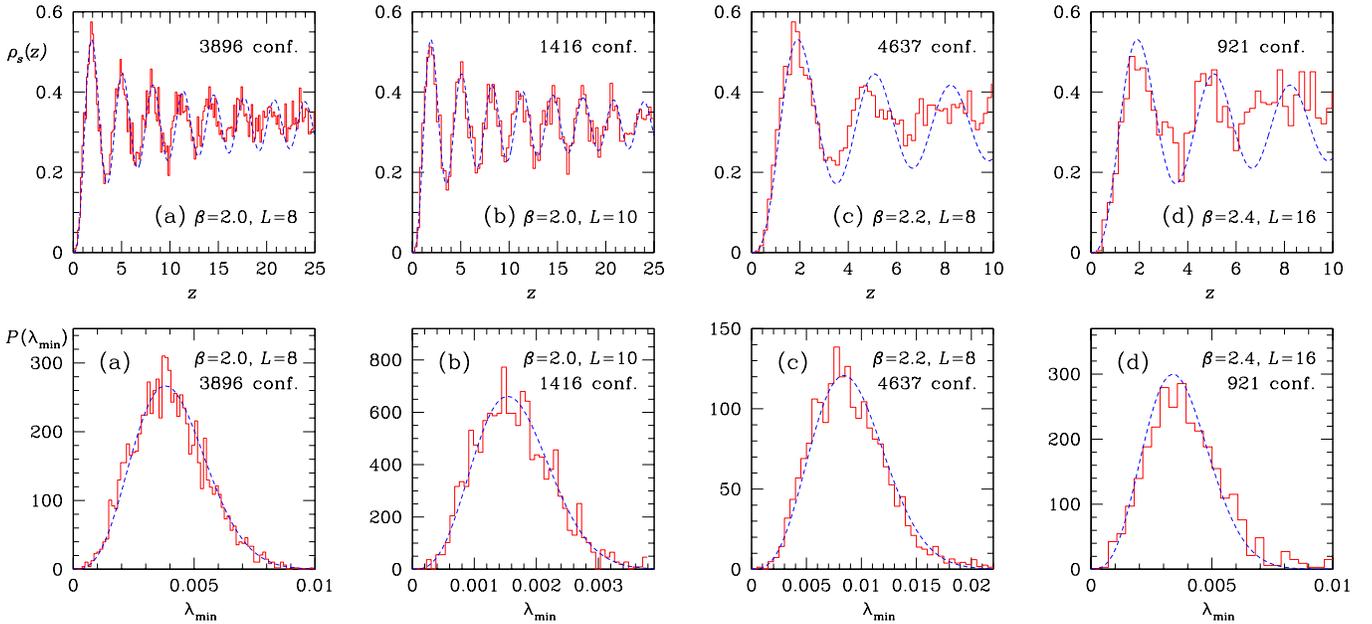,width=\textwidth}}
\widetext
\vspace*{2mm}
\caption{Microscopic spectral density (upper row) and distribution of
  the smallest eigenvalue of the Dirac operator for different lattice
  parameters. From left to right the values of $\beta$ are: 2.0, 2.0,
  2.2, and 2.4.  The histograms represent lattice data, the dashed
  curves are predictions from chiral RMT with $N_f=\nu=0$.}
\vspace*{2mm}
\label{fig2}]
\narrowtext
\end{figure}
In Fig.~\ref{fig2} we have plotted the lattice results for $\rho_s(z)$
and $P(\lambda_{\rm min})$ together with the analytical results of
Eqs.~(\ref{eq3}) and (\ref{eq4}) for four different combinations of
$\beta$ and lattice size.  The agreement between lattice data and the
parameter-free RMT predictions is impressive.  Note that the RMT
results were derived in the limit $V\to\infty$. Clearly, the agreement
improves as the physical volume increases, i.e., with larger lattice
size and smaller $\beta$. From the results for $\beta =2.0$ we observe
that the agreement with RMT improves with increasing lattice size
while the value of condensate remains the same. This suggests that a
similar improvement will occur for $\beta =2.2$ and $\beta = 2.4$.
These values are just below the $\beta$-value above which $\langle
\bar \psi \psi \rangle$ approaches zero, where the above RMT results
are inapplicable, and an increased sensitivity to the size of the
lattice is expected.  Since $P(\lambda_{\rm min})$ for these couplings
agrees with the RMT distribution for zero topological charge we expect
that the discrepancy for $\rho_s(z)$ is not due to a superposition of
configurations with different topological charge. We hope that future
work will clarify this issue.

\begin{figure}[ht]
\centerline{\psfig{figure=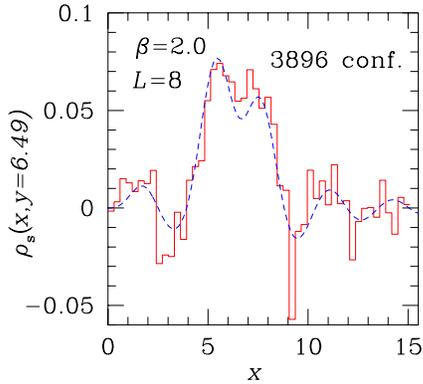,height=50mm}}
\vspace*{2mm}
\caption{Microscopic limit of the two-level cluster function for some
  fixed value of $y$.  The histogram represents data for $\beta=2.0$
  on an $8^4$ lattice, the dashed curve shows the random-matrix
  prediction.}
\label{fig3}
\end{figure}

Related quantities testing similar properties are the higher-order
spectral correlation functions, in particular the two-point function
which enters in the computation of scalar susceptibilities.  The
$n$-point correlation function $R_n(x_1,\ldots,x_n)$ is defined as the
probability density of finding a level (regardless of labeling) around
each of the points $x_1,\ldots,x_n$.  The two-level cluster function
$T_2(x,y)$, which contains only the non-trivial correlations, is
defined by $T_2(x,y)=-R_2(x,y)+R_1(x)R_1(y)$, i.e., the disconnected
part is subtracted.  For the chGUE, there are analytical arguments
\cite{Guhr97} that the microscopic correlations are universal, and the
same is expected for the chGSE.  In this case, the predictions from
RMT can again be obtained from the results of Ref.~\cite{Naga95}, but
we do not write down the explicit expressions here.  In
Fig.~\ref{fig3}, we have plotted data for $\rho_s(x,y)$ for
$\beta=2.0$ on an $8^4$ lattice as a function of $x$ for some fixed
value of $y$, along with the analytical random-matrix prediction.
Clearly, the statistics are not as good as for the one-point function,
but the agreement is still quite impressive.

Finally, we have checked the Leutwyler-Smilga sum rule
$\langle\sum_n\lambda_n^{-2}\rangle/V^2=\langle\bar\psi\psi\rangle^2/2$
appropriate for the chGSE \cite{Leut92,Verb94c}.  The numerical
results are compared with the analytical predictions in
Table~\ref{table1}.  Again, the agreement improves with physical
volume.
\begin{table}[hb]
\caption{Chiral condensate and a comparison of lattice data and
  analytical predictions for the Leutwyler-Smilga sum rule for
  $\lambda_n^{-2}$.}
\begin{tabular}{ccccc}
  $\beta$ & $L$ & $\langle\bar\psi\psi\rangle$ &
  $\langle\sum_n\lambda_n^{-2}\rangle/V^2$ &
  $\langle\bar\psi\psi\rangle^2/2$ \\[0.5mm]
  \tableline
  2.0 & 8  & 0.1228(25) & $8.20(20)\!\cdot\!10^{-3}$ & 
  $7.54(31)\!\cdot\!10^{-3}$ \\
  2.0 & 10 & 0.1247(22) & $7.97(30)\!\cdot\!10^{-3}$ & 
  $7.78(27)\!\cdot\!10^{-3}$ \\
  2.2 & 8  & 0.0556(19) & $1.67(03)\!\cdot\!10^{-3}$ & 
  $1.55(11)\!\cdot\!10^{-3}$ \\
  2.4 & 16 & 0.00863(48) & $3.97(14)\!\cdot\!10^{-5}$ & 
  $3.72(42)\!\cdot\!10^{-5}$ 
\end{tabular}
\label{table1}
\end{table}

In summary, we have performed a high-statistics study of the
eigenvalue spectrum of the lattice QCD Dirac operator with particular
emphasis on the low-lying eigenvalues.  In the absence of a formal
proof, our results provide very strong and direct evidence for the
universality of $\rho_s$. In the strong coupling domain, the agreement
with analytical predictions from random matrix theory is very good.
On the scale of the smallest eigenvalues, agreement is found even in
the weak-coupling regime. Furthermore, we found that the microscopic
two-level cluster function agrees nicely with random-matrix
predictions and that the Leutwyler-Smilga sum rule for
$\lambda_n^{-2}$ is satisfied more accurately with increasing physical
volume.  We predict that corresponding lattice data for SU(2) with
Wilson fermions and for SU(3) with staggered and Wilson fermions will
be described by random matrix results for the GOE, chGUE, and GUE,
respectively \cite{Verb94a}.  (The U$_{\rm A}$(1) symmetry is absent
for the Hermitean Wilson Dirac operator.)  The identification of
universal features in lattice data is both of conceptual interest and
of practical use.  In particular, the availability of analytical
results allows for reliable extrapolations to the chiral and
thermodynamic limits.  In future work we hope to analyze the fate of
the fermionic zero modes in the approach to the continuum limit, and
we expect random-matrix results to be a useful tool in the analysis.

It is a pleasure to thank T.\ Guhr and H.A.\ Weidenm\"uller for
stimulating discussions.  This work was supported in part by DFG and
BMBF.  SM and AS thank the MPI f\"ur Kernphysik, Heidelberg, for
hospitality and support.  The numerical simulations were performed on
a CRAY T90 at the Forschungszentrum J\"ulich and on a CRAY T3E at the
HWW Stuttgart.

\end{document}